\documentclass[twocolumn,superscriptaddress,floatfix,preprintnumbers, nofootinbib,hyperref]{revtex4} 
\pdfoutput=1
\usepackage[colorlinks=true,breaklinks=true]{hyperref}
\usepackage[normalem]{ulem}
\usepackage[utf8]{inputenc}
\hypersetup{allcolors=[rgb]{0.0 0.0 0.6},linkcolor=[rgb]{0.75 0.05 0.05}}
\usepackage{amsmath,amssymb}
\usepackage{epsfig}  
\usepackage{graphicx}   
\usepackage{slashed}             
\usepackage{url}
\usepackage{color}
\usepackage{multirow}
\usepackage{subfigure}
\usepackage{comment}
\usepackage[dvipsnames]{xcolor}
\usepackage{letltxmacro}
\LetLtxMacro{\oldcite}{\cite}
\renewcommand{\cite}[1]{\mbox{\oldcite{#1}}}

\DeclareMathOperator{\MeV}{MeV}

\DeclareMathOperator{\kpc}{kpc}

\newcommand{\beq}{\begin{equation}}
\newcommand{\eeq}{\end{equation}}

\allowdisplaybreaks

\bibpunct{[}{]}{,}{n}{}{,}

\definecolor{darkblue}{rgb}{1, 0.1, 0.2}
\begin{document}

\title{Supernova bounds on axion-like particles coupled with nucleons and  electrons}

\author{Francesca Calore}
 \email{calore@lapth.cnrs.fr}
\affiliation{Univ.~Grenoble Alpes, USMB, CNRS, LAPTh, F-74000 Annecy, France}

\author{Pierluca Carenza}
\email{pierluca.carenza@ba.infn.it }
\affiliation{Dipartimento Interateneo di Fisica ``Michelangelo Merlin'', Via Amendola 173, 70126 Bari, Italy.}
\affiliation{Istituto Nazionale di Fisica Nucleare - Sezione di Bari,
Via Orabona 4, 70126 Bari, Italy.}

\author{Maurizio Giannotti}
\email{MGiannotti@barry.edu}
\affiliation{Physical Sciences, Barry University, 11300 NE 2nd Ave., Miami Shores, FL 33161, USA}

\author{Joerg Jaeckel}
\email{jjaeckel@thphys.uni-heidelberg.de}
\affiliation{Institut f\"ur theoretische Physik, Universit\"at Heidelberg,
Philosophenweg 16, 69120 Heidelberg, Germany}

\author{Giuseppe Lucente}
\email{giuseppe.lucente@ba.infn.it }
\affiliation{Dipartimento Interateneo di Fisica ``Michelangelo Merlin'', Via Amendola 173, 70126 Bari, Italy.}
\affiliation{Istituto Nazionale di Fisica Nucleare - Sezione di Bari,
Via Orabona 4, 70126 Bari, Italy.}

\author{Alessandro Mirizzi}
\email{alessandro.mirizzi@ba.infn.it }
\affiliation{Dipartimento Interateneo di Fisica ``Michelangelo Merlin'', Via Amendola 173, 70126 Bari, Italy.}
\affiliation{Istituto Nazionale di Fisica Nucleare - Sezione di Bari,
Via Orabona 4, 70126 Bari, Italy.}

%=============================================================================

\begin{abstract}
We investigate the potential of type II supernovae (SNe) to constrain axion-like particles (ALPs) coupled simultaneously to nucleons and electrons. ALPs coupled to nucleons can be efficiently produced in the SN core via nucleon-nucleon bremsstrahlung and, for a wide range of parameters, leave the SN unhindered, producing a large ALP flux.
For masses exceeding 1 MeV, these ALPs would decay into electron-positron pairs, generating a positron flux. In the case of Galactic SNe, the annihilation of the created positrons with the electrons present in the Galaxy would contribute to the 511 keV annihilation line. Using the SPI (SPectrometer on INTEGRAL) observation of this line, allows us to exclude a wide range of the axion-electron coupling, $10^{-19} \lesssim g_{ae} \lesssim 10^{-11}$, for $g_{ap}\sim 10^{-9}$. Additionally, ALPs from extra-galactic SNe decaying into electron-positron pairs would yield a contribution to the cosmic X-ray background. In this case, we constrain the ALP-electron coupling down to $g_{ae} \sim 10^{-20}$.
\end{abstract}

\maketitle

\section{Introduction}

Featuring large densities and core temperatures $T \sim \mathcal{O}(30)$~MeV, type II supernovae (SNe) allow for  the prolific production of ALPs with masses up to $100$~MeV~\cite{Raffelt:1996wa}. In this paper we complement our previous study~\cite{Calore:2020tjw} of SNe constraints on combinations of the ALP-nucleon and ALP-photon couplings by deriving limits on ALPs coupled to both nucleons and electrons.

There are two types of arguments typically used to constrain ALPs from SNe. The \emph{first strategy} is based on an \emph{indirect} signature, due to the impact of the ALP emission on the SN neutrino signal: an overly efficient ALP production during the early stage of the SN evolution would significantly shorten the duration of the SN neutrino burst~\cite{Raffelt:2006cw}. The time distribution of the few neutrinos observed from SN 1987A thus allows to exclude sizable exotic energy losses. 

In the case of QCD axions or generic ALPs coupled with \emph{nucleons}, the most efficient emission process in a SN core is nucleon-nucleon bremsstrahlung~\cite{Raffelt:2006cw,Burrows:1988ah}. In this case, the neutrino signal argument provides the stringent bound $g_{aN} \lesssim 10^{-9}$~\cite{Burrows:1988ah,Keil:1996ju,Raffelt:2006cw,Carenza:2019pxu}\footnote{It has been recently pointed out~\cite{Carenza:2020cis} that if the SN core contains a significant fraction of thermal pions the previous bound may be strengthened by a factor $\sim 2$. However, since the impact of pions on the ALP emissivity is still the object of current investigations, we will not use this argument here. We will also only consider the mass region where the effects of the ALP mass on the flux is quite small, $m_{a}\lesssim 30$~MeV. For larger masses an improved calculation for the bremsstrahlung to massive ALPs would be needed (for an estimate using phase space arguments see~\cite{Carmona:2021seb})} on the ALP-nucleon coupling\footnote{For general arguments we will typically refer to the ALP nucleon coupling $g_{aN}$ without precisely specifying whether it is the proton or the neutron coupling (or even a combination). However, for the discussion of the limits we will be more concrete and specify that our numbers are obtained with a given value of $g_{ap}$ and $g_{an}=0.$}.

ALPs can be produced in the SN core also through couplings to other fields. If coupled to photons, ALPs can be created through the Primakoff process, i.e. the conversion of thermal photons in the electric field of protons. However, this mechanism is not very efficient and the constraint derived from the neutrino signal is considerably less stringent than the upper bound derived from the evolution of low mass stars~\cite{Ayala:2014pea,Giannotti:2015kwo,Giannotti:2017hny}. An exception is the case of ALPs with masses $\gtrsim 100\,{\rm keV}$~\cite{Lucente:2020whw}, since such heavy particles cannot be efficiently produced in low temperature stars.

The ALP coupling to electrons may also lead to a production of ALPs in the SN core, mostly through electron bremsstrahlung or, for $m_{a} \gtrsim 10 \MeV$, electron-positron annihilation $e^{+}e^{-}\rightarrow a$, leading to a bound on the ALP-electron coupling $4\times10^{-10}<g_{ae} <5\times10^{-8}$ for $m_{a}<300\MeV$~\cite{Lucente:2021hbp}. 

The \emph{second strategy} to constrain the ALP couplings from SNe, which is the one we will employ in this work, relies instead on the search for \emph{direct} signatures of the SN ALP flux. A well studied example is that of light ALPs ($m_a \lesssim 10^{-10}$~eV) coupled to photons, produced in the SN core through the Primakoff process and converted into gamma rays in the magnetic field of the Milky Way~\cite{Grifols:1996id,Brockway:1996yr}.
This mechanism predicts a well defined and in principle observable gamma-ray signal. The non-observation of such a signal in the Gamma-Ray Spectrometer (GRS) on the Solar Maximum Mission (SMM) in coincidence with the observation of the neutrinos emitted from SN 1987A allows to set a strong bound on ALPs coupling to photons~\cite{Grifols:1996id,Brockway:1996yr}. The most recent analysis finds $g_{a\gamma} < 5.3 \times 10^{-12}$ GeV$^{-1}$ for $m_a < 4 \times 10^{-10}$~eV~\cite{Payez:2014xsa}. 

Heavy ALPs, with mass $m_a \sim {\mathcal O}(0.1-100)$~MeV and coupled to photons, would produce a gamma-ray signal through the decay $a\to \gamma \gamma$, rather than conversion~\cite{Giannotti:2010ty,Jaeckel:2017tud}. In this case, the non-observation of a gamma-ray signal in coincidence with the SN 1987A implies the bound $g_{a\gamma} \lesssim   10^{-11}$ GeV$^{-1}$ at $m_{a}\sim 10$~MeV~\cite{Jaeckel:2017tud}. Intriguing opportunities to sharpen these bounds are offered by the detection of an ALP burst in future (extra)-galactic SN explosions~\cite{Meyer:2016wrm,Meyer:2020vzy} or from the analysis of the diffuse ALP flux from all past type II SNe in the Universe~\cite{Raffelt:2011ft,Calore:2020tjw}. 
 
\bigskip
 
From a phenomenological point of view, it is interesting to consider ALPs simultaneously coupled to several Standard Model (SM) fields. In fact, generically axion and ALP models feature more than one non-vanishing coupling~\cite{DiLuzio:2020wdo,Agrawal:2021dbo}.

Combinations of ($g_{a\gamma}$-$g_{ae}$) have been studied in the case of the Sun~\cite{Barth:2013sma,Jaeckel:2018mbn,Hoof:2021mld} and globular cluster stars~\cite{Giannotti:2015kwo,Giannotti:2017hny}. As we will show, SNe are also sensitive tools to probe different combinations of couplings of ALPs with SM particles. Our recent investigation, in Ref.~\cite{Calore:2020tjw}, considered ALPs simultaneously coupled to photons and nucleons (see also~\cite{Giannotti:2010ty}). In this case, the nucleon coupling would be responsible for the ALP production through nuclear bremsstrahlung, a mechanism considerably more efficient than the Primakoff process. Assuming a nucleon coupling roughly equal to the bound from SN 1987A, the limit on the ALP-photon coupling for ultralight ALPs would be pushed down to $g_{a\gamma} < 6\times 10^{-13}$ GeV$^{-1}$, while in the massive case it was found to be $g_{a\gamma} \lesssim 10^{-19}$ GeV$^{-1}$ at $m_{a}\sim 20$~MeV.

Given these results, we find it useful to investigate also combinations of the ALP-nucleon coupling $g_{aN}$ with the ALP-electron coupling $g_{ae}$.\footnote{See Ref.~\cite{Giannotti:2010ty} for some early comments in this direction.}
In this work, we consider ALP masses $m_a \gtrsim 1$ MeV, to allow the ALP decay into electron-positron pairs. If the coupling with electrons is sufficiently small, ALPs would  leave the SN envelope and decay on their route to Earth. The positrons produced in these decays are expected to efficiently lose energy, slow down, and annihilate almost at rest with the Galactic electron density, leading to a characteristic 511 keV annihilation line signal. 
This line has been well measured by SPI (SPectrometer on INTEGRAL)~\cite{Strong:2005zx,Bouchet:2010dj,Siegert:2015knp, Siegert:2019tus}.~\footnote{An additional measurement of the 511 keV line has been recently performed by the Compton Spectrometer and Imager (COSI) experiment~\cite{Kierans:2019aqz}. Since the results are comparable with those of SPI, we will not make use of them in the following.} In the following, we will constrain the ALP-electron and ALP-nucleon coupling using observations of the 511 keV line flux, and exploiting its spatial characterization.

Following a phenomenological approach, we mostly consider ALPs that are coupled dominantly to electrons and nuclei with a negligible coupling to photons (see~\cite{Craig:2018kne} for ``photophobic'' scenarios). However, as we will briefly discuss in Sec.~\ref{sec:production}, our considerations {are valid} even for photon couplings of a size expected for typical pseudo-Goldstone bosons.

The plan of our work is the following: in Sec.~\ref{sec:production} we recall the ALP flux from nucleon-nucleon bremsstrahlung and discuss the decay rate into electron-positron pairs for massive ALPs. Using the decay of ALPs into electron-positron pairs, in Sec.~\ref{sec:positron} we derive our bounds on the ALP couplings. In particular, in Sec.~\ref{sec:galactic} we consider Galactic SNe and derive the bound from the 511 keV signal, while in Sec.~\ref{sec:extra-galactic} we  explore the case of ALP decay from extra-galactic SNe, with the positrons annihilating outside the Galaxy. In this last case, the resulting photon flux, properly redshifted, contributes to the cosmic X-ray diffuse background, allowing us to place an additional bound. In Sec.~\ref{sec:conclusions}, we comment on our results and conclude. Finally, in the Appendix we discuss the possible uncertainties affecting our results.

\section{ALPs from Supernovae: production and decays}\label{sec:production}

\subsection{ALP production in a SN core}

The ALP interactions with the SM fields are expressed by the following Lagrangian terms~\cite{DiLuzio:2020wdo}
\begin{equation}
{\mathcal L}_{\rm int}=  
 \sum_{\psi=e,p,n}\frac{ g_{a\psi}}{2 m_\psi} ( \bar{\psi}\gamma_\mu \gamma_5 \psi) \partial^\mu a 
 -\frac{1}{4} \,g_{a\gamma}
F_{\mu\nu}\tilde{F}^{\mu\nu}a\,\ ,
\label{Eq:interactions}
\end{equation}
where $g_{a\psi}$ are the effective (dimensionless) ALP couplings with fermions with mass $m_\psi$, and $g_{a\gamma}$ is the photon-ALP coupling constant (with dimension of an inverse energy).

In the following, we assume the coupling to photons to be small enough to guarantee a much more efficient ALP decay into electron-positron pairs than into photons (cf.~Sec.~\ref{sec:ALP_decay}). With this assumption, the ALP production rate in the SN proceeds mostly through bremsstrahlung, $\psi+\psi\to \psi+\psi + a$, where $\psi$ is any of the fermions in Eq.~\eqref{Eq:interactions}. However, in the range of parameters we are exploring in this work, the contribution of the electron bremsstrahlung to the SN ALP production can be ignored and the nuclear bremsstrahlung remains  the only significant production mechanism. A recent evaluation of the ALP flux generated by the nucleon bremsstrahlung process can be found in Ref.~\cite{Carenza:2019pxu,Calore:2020tjw}, to which we refer the interested reader for further details. For the purpose of our work here, it is sufficient to point out that, in the case of $m_a \ll T$, the integrated ALP spectrum is given, to excellent precision, by the analytical expression~\cite{Calore:2020tjw}
\begin{equation}
\frac{dN^{p}_a}{dE} = C \left(\frac{g_{ap}}{g^{\rm ref}_{ap}}\right)^2
\left(\frac{E}{E_0}\right)^\beta \exp\left( -\frac{(\beta + 1) E}{E_0}\right) \,,
\label{eq:time-int-spec}
\end{equation}
where $C=9.08\times 10^{55}$ ${\rm MeV}^{-1}$, $E_0=103.2$ MeV, and $\beta=2.2$ for the reference couplings $g^{\rm ref}_{ap}=10^{-9}$ and $g_{an}=0$. These values are obtained from a SN model with an 18~$M_{\odot}$ progenitor, simulated in spherical symmetry with the AGILE-BOLTZTRAN code~\cite{Mezzacappa:1993gn,Liebendoerfer:2002xn}. This is a reasonable approximation for the ALP spectrum for masses $m_{a}\lesssim 30\,{\rm MeV}$, including accurate nuclear physics and many-body effects~\cite{Carenza:2019pxu}. In what follows, we will assume this SN as representative of all type II SN models. A discussion of the uncertainty introduced by this assumption is presented in the Appendix, where we evaluate this bound assuming a $11.2\,$M$_{\odot}$ and a $25\,$M$_{\odot}$ SN as the representative model. A more in depth discussion of this, including the mass distribution of SNe, is foreseen for future work~\cite{future}.

To ensure that a putative photon signal is not absorbed by the SN medium, we require that ALPs decay outside the SN envelope, which we take to have a size $r_{\rm esc}=10^{14}$~cm~\cite{DeRocco:2019njg}. This requires $g_{ae}\lesssim 10^{-11}-10^{-12}$, somewhat depending on the ALP mass (see also Eq.~\eqref{eq:decayel} below). 

\subsection{ALP decays into electron-positron pairs}
\label{sec:ALP_decay}

ALPs in the mass range, $1\,{\rm MeV}<m_{a}\lesssim 100\,{\rm MeV}$ can decay only into photons and electron-positron pairs. The partial decay lengths are given by (see Ref.~\cite{Altmann:1995bw,Jaeckel:2017tud})
\begin{equation}
\begin{split}
l_{e}&=\frac{\gamma v}{\Gamma_{a\rightarrow e^{+}e^{-}}}= 
\frac{E_a}{m_a}\frac{\sqrt{1-\frac{m_a^2}{E_a^2}}}{\sqrt{1-\frac{4 m_e^2}{m_a^2}}} \frac{8 \pi}{g_{ae}^2 m_a} 
\\
&\simeq \!1.6\times10^{-5}\,\textrm{kpc}\! \left(\frac{E_a}{100 ~\textrm{MeV}}\right)\! \left(\frac{10~\textrm{MeV}}{m_a} \right)^2\!\left(\frac{10^{-13}}{g_{ae}}\right)^2 \! ,
\end{split}
\label{eq:decayel}
\end{equation}
and
\begin{equation}
\begin{split}
l_{\gamma}&=\frac{\gamma v}{\Gamma_{a\rightarrow\gamma\gamma}}= 
\frac{E_a}{m_a}\sqrt{1-\frac{m_a^2}{E_a^2}}  \frac{64 \pi}{g_{a\gamma}^2 m_a^3} 
\\
&\simeq \! 1.3~\textrm{kpc}\! \left(\frac{E_a}{100 ~\textrm{MeV}} \right)\! \left(\frac{10~\textrm{MeV}}{m_a} \right)^4\!\left(\frac{10^{-13} \textrm{GeV}^{-1}}{g_{a\gamma}} \right)^2 \! .
\end{split}
\label{eq:decaygamma}
\end{equation}
The total ALP decay length is $l_{\rm tot}^{-1}=l_{e}^{-1}+l_{\gamma}^{-1}$. In the present analysis, we focus mostly on the case in which the contribution of $l_\gamma$ to the decay length is negligible. For this to be a reasonable assumption, the branching ratio into electrons must be dominant, i.e.
\begin{equation}
\begin{split}
&\frac{BR(a\to \gamma\gamma)}{BR(a\to e^{+}e^{-})}
= \frac{l_e}{l_\gamma} =\\
&\sim 10^{-5} \left(\frac{m_a} {10~ \textrm{MeV}}\right)^2\left(\frac{10^{-13}}{g_{ae}}\right)^{2} \left(\frac{g_{a\gamma}}{10^{-13}~ \textrm{GeV}^{-1}} \right)^2 \ll 1.
\end{split}
\label{eq:lengthrelation}
\end{equation}

For a typical pseudo-Goldstone where one universal decay constant $f$ sets all relevant couplings we expect the relations,
\begin{eqnarray}
\label{eq:natural}
g_{ae}&\sim& \frac{m_{e}}{f}\sim 10^{-13}\left(\frac{10^{10}\,{\rm GeV}}{f}\right)
\\\nonumber
g_{a\gamma}&\sim& \frac{\alpha}{4\pi f} \sim 10^{-13}\,{\rm GeV}^{-1} \left(\frac{10^{10}\,{\rm GeV}}{f}\right)
\\\nonumber
g_{aN}&\sim&  {\mathcal{O}}(1)\times \frac{m_{N}}{f} \sim 10^{-10} \left(\frac{10^{10}\,{\rm GeV}}{f}\right),
\end{eqnarray}
where in the last line the ${\mathcal{O}}(1)$ constant depends on whether the underlying couplings are to quarks or gluons and whether we are dealing with a proton or a neutron (see, e.g.~\cite{diCortona:2015ldu,Irastorza:2018dyq,DiLuzio:2020wdo} for values in the specific case of QCD axions).

From the relation for the electron and the photon coupling we can see that for masses $m_a \lesssim 100\,{\rm MeV}$ the requirement that the decay occurs dominantly into electrons, Eq.~\eqref{eq:lengthrelation}, is naturally fulfilled in the case of a universal decay constant. A similar parametric relation can be obtained by considering only an electron coupling that then generates a photon coupling via a loop diagram, cf. the expressions in~\cite{Bauer:2017ris,Craig:2018kne}.

In what follows, we will also set constraints on electron couplings that are considerably suppressed compared to the nucleon couplings. While our approach is agnostic about the origin of the couplings, let us mention that such couplings could occur, for example, in a photophobic scenario~\cite{Craig:2018kne} where, in addition, the electron coupling is loop suppressed. In the case of suppressed electron couplings, we then have to check that the loop generated photon coupling from the nucleon coupling are not too large, and the decay into electrons is still dominant. Naively applying the loop expressions\footnote{In principle some care might be necessary to take the bound state nature of the nucleons into account.} from~\cite{Bauer:2017ris,Craig:2018kne} this is of the order,
\begin{eqnarray}
g_{a\gamma}\!\!&\sim&\!\! \frac{\alpha}{6\pi}\frac{m^{2}_{a}}{m^{3}_{N}}g_{aN}
\\\nonumber
\!\!&\sim&\!\! 4\times 10^{-18}{\rm GeV}^{-1}\,\left(\frac{m_{a}}{10\,{\rm MeV}}\right)^2\left(\frac{g_{aN}}{10^{-10}}\right).
\end{eqnarray}
For a wide range of masses and electron couplings, this is compatible with a dominant decay into electrons according to Eq.~\eqref{eq:lengthrelation}. However, for the smallest values of $g_{ae}$ and higher masses $m_{a}$, some tuning may be required to eliminate this coupling.

\section{Positron bounds}\label{sec:positron}

As discussed above, we are interested in ALPs that decay dominantly into electron-positron pairs, outside the SN envelope. The generated positrons are then trapped by the $\mathcal{O}(1)~\mu {\rm G}$ Galactic magnetic field, and lose their energy efficiently through Bhabha $e^+ e^-$ scatterings, before annihilating (almost at rest) into two photons, each with an energy of  $\sim 511$~keV. Strictly speaking, a magnetic field traps charged particles only in the direction perpendicular to the magnetic field. Nevertheless, for typical conditions of the interstellar medium, positrons with energies $\lesssim 100$~MeV are expected to travel not more than $\sim1$~kpc~\cite{Jean:2005af,Jean:2009zj,Martin:2012hv}. In the following we will take a distance of $1$~kpc as our baseline propagation length. 

Depending on the free electron density in our Galaxy and the ionization conditions of the inter-stellar medium, the positron annihilation time is expected to range between $\tau_e \in \left[10^{3}-10^{6}\right]$ years \cite{Wang:2005cqa,Kalemci:2006bz}. We neglect the so-called in-flight annihilation channel of high-energy positrons, unless otherwise stated~\cite{Beacom:2005qv}. While this could, in principle, reduce the fraction of positrons  being stopped and contributing to the 511 keV line, the signal reduction is not expected to exceed 25\%. On the other hand photons produced by the in-flight annihilation would contribute to the diffuse MeV-GeV emission and, therefore, offer another possible mean to set constraints on our model.

The SPI gamma-ray spectrometer on the INTEGRAL satellite provides measurements of the Galactic 511 keV X-ray line flux~\cite{Strong:2005zx,Bouchet:2010dj,Siegert:2015knp, Siegert:2019tus}. We will use observations from~\cite{Siegert:2019tus} to constrain the positron flux injected by ALPs produced by Galactic SNe and decaying outside the SN envelope. This allows us to obtain a bound on $g_{ae}$ and $g_{aN}$. Additionally, we will use measurements of the cosmic X-ray background (CXB) by the High Energy Astronomy Observatory  (HEAO)~\cite{McHardy:1997fb} and the Solar Maximum Mission (SMM)~\cite{doi:10.1063/1.53933} to constrain a diffuse flux generated by all the past extra-galactic SNe.

\subsection{Flux from Galactic supernovae}
\label{sec:galactic}

The SPI gamma-ray spectrometer measurements constrains the Galactic center positron annihilation rate to be smaller than a few $\times 10^{43}$~s$^{-1}$~\cite{Prantzos:2010wi,Raffelt:1996wa}. Since electron-positron annihilation seems to be in equilibrium, this can also be taken as a bound on the positron production rate. 
Assuming a Galactic SN rate of 2 events per century, in Ref.~\cite{DeRocco:2019njg} it was estimated that the previous bound would be saturated if a single SN emits more than $10^{53}$ positrons. This result was then used, in the same work, to obtain constraints on dark photons emitted from SNe and decaying into electron-positron pairs. This argument is, however, somewhat oversimplified since it does not take into account the specific distribution of the positrons in the Galaxy, assuming instead that it has the same morphology of the detected 511 keV line. 
As we will see, this is not fully realistic.

In our work, we adopt a more detailed approach, exploiting the longitude and latitude  distributions of the 511 keV gamma-ray flux provided by an analysis of SPI data~\cite{Siegert:2019tus}, and comparing it with the expected 511 keV line signal produced by positrons from ALP decays, as traced by the probability distribution of type II SNe. 

The time integrated flux of ALPs, produced in the SN core and  escaping from the SN envelope is given by 
\begin{eqnarray}
\left(\frac{dN_{\rm a}^p}{dE}\right)_{\rm esc}&=&\frac{dN_{\rm a}^p}{dE} \times\exp \bigg(-\frac{r_{\rm esc}}{l_{e }}\bigg)
\end{eqnarray}
where the ALP production rate, $dN_{\rm a}^p/dE$, is given by Eq.~(\ref{eq:time-int-spec}). For simplicity we assume that all SNe are well represented by our model and neglect the sub-leading contribution of type Ib/c SNe to the SN rate. Then, the injected positron flux from a SN explosion is given by
\begin{equation}
\label{eq:positrons}
N_{\rm pos}=\int dE \left(\frac{dN_{\rm a}^p}{dE}\right)_{\rm esc}\bigg[1 - \exp\bigg(-\frac{r_{\rm G}}{l_{e}}\bigg) \bigg] \,\ , 
\end{equation}
where the term in square brackets guarantees that the ALPs decay within our Galaxy. To be conservative, we use a very small value for the radius, $r_G=1$ kpc, which ensures that we stay inside the Galaxy in all directions (including, in particular, the direction perpendicular to the plane). 

\begin{figure}[t!]
\vspace{0.cm}
\includegraphics[width=0.9\columnwidth]{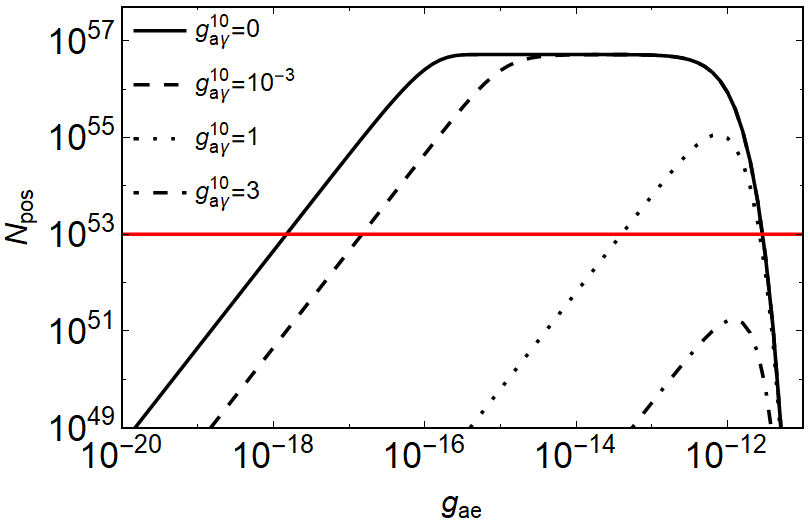}
\caption{Number of positrons produced inside a radius of $r_G=1$~kpc, per SN as function of $g_{ae}$ for different values of $g_{a\gamma}^{10}\equiv g_{a\gamma}/10^{-10}{\rm GeV}^{-1}$, and a fixed $g_{ap}=10^{-9}$ and $m_{a}=30\,\MeV$. The red line indicates a sizable positron production, $N_{\rm pos}=10^{53}$~\cite{DeRocco:2019njg}.
}
\label{fig:positrons}
\vspace{0.3cm}\includegraphics[width=0.95\columnwidth]{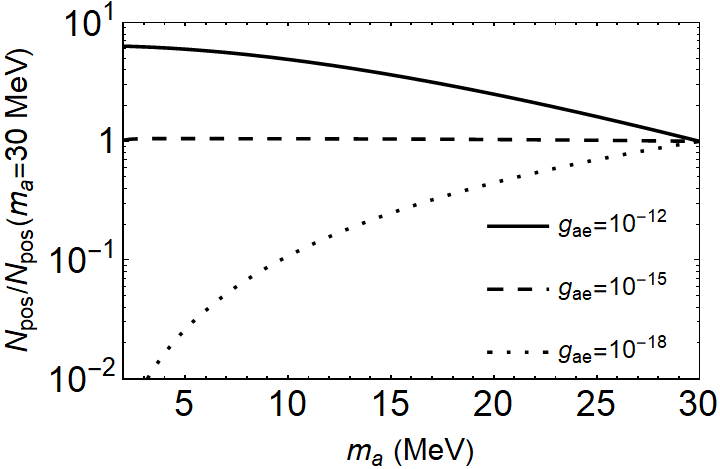}
\caption{Number of positrons as function of the ALP mass for $g_{a\gamma}=0$ and different values of the ALP-electron coupling.
}
\label{fig:Nposvsma}
\end{figure}

In Fig.~\ref{fig:positrons}, we show the number of positrons that are produced \emph{inside} the Galaxy (black solid line). As a reference, the horizontal line indicates the maximal positron number ($N_{\rm pos}=10^{53})$ used in Ref.~\cite{DeRocco:2019njg}.

For a vanishing ALP-photon coupling and a small $g_{ae}$, the positron production is suppressed since most ALPs escape the Galaxy before decaying. As $g_{ae}$ increases, ALPs decay inside the Galaxy, reaching the maximum positron production when all the ALPs decay in the Galaxy, for $g_{ae}\gtrsim 10^{-16}$. This plateau extends for some orders of magnitude in $g_{ae}$, depending on the ALP mass (e.g. up to $g_{ae}\simeq 10^{-12}$ for  $m_{a}=30\,{\rm MeV}$). For larger $g_{ae}$, ALPs decay inside the SN envelope, and the created positrons annihilate in an environment where it is not clear whether they can escape and contribute to an observable signal.

At this point let us comment on the effects of a non-vanishing coupling $g_{a\gamma}$. Its effect is demonstrated in the non-solid lines of Fig.~\ref{fig:positrons}. In order to take into account the impact of decays into photons on the positron flux we multiply  Eq.~\eqref{eq:positrons} by the branching ratio 
\begin{equation}
\frac{BR(a\to e^+e^-)}
{BR(a\to \gamma \gamma)+
BR(a\to e^+e^-)} \,\ .    
\end{equation}
The effect of the photon coupling in reducing the branching ratio is particularly important for small $g_{ae}$, where the escape-related suppression factor is linear in $l_{\rm tot}^{-1}$. 

In Fig.~\ref{fig:Nposvsma}, we show the number of positrons produced, $N_{\rm pos}$, as a function of the ALP mass $m_a$, for three different values of $g_{ae}$ and keeping $g_{a\gamma}=0$. As a benchmark value we use $m_a=30$~MeV. For $g_{ae}=10^{-12}$ (solid curve) the positron number monotonically decreases while increasing the ALP mass. The behavior is due to the fact that, for such large ALP-electron couplings, a large portion of ALPs decay before leaving the SN and thus do not contribute to the positron flux. This fraction clearly increases at higher masses since in this case the decay length is reduced, as evident from Eq.~(\ref{eq:decayel}). For $g_{ae}=10^{-15}$ (dashed curve), the positron number is rather insensitive to the ALP mass. This case corresponds to a regime in which all ALPs decay outside the SN and inside the Galaxy, giving the maximum contribution to $N_{\rm pos}$. Finally, for $g_{ae}=10^{-18}$, $N_{\rm pos}$ increases as function of the ALP mass.  
Indeed, for such a small value of the ALP-electron coupling, a reduction in the decay length would enhance the ALPs probability of decaying inside the Galaxy.

\bigskip

Let us now consider the distribution of the generated photons and compare it to the data. In addition to the location of the SNe producing the ALPs there are two factors that determine the morphology of the 511 keV signal. First, depending on their decay length ALPs may travel a considerable distance before decaying. To be conservative we only count positrons from ALP decays within 1~kpc such that we are safely within the galaxy. Second, the positrons produced from the ALP decays may travel some distance before being stopped and annihilated. As long as the SN ALPs decay in the Galaxy, the 511 keV photons are expected to be generated not farther than $1$~kpc~\cite{Jean:2005af,Jean:2009zj,Martin:2012hv} from the ALP decay region. We estimate the influence of both effects by performing a smearing of the signal on a 1~kpc scale as described below. In the Appendix we show the variation of the bound related to a different choice of the smearing scale.

Let us also briefly comment on the time structure of the signal. Most positrons with energies $\lesssim 100$~MeV, slow down before undergoing an almost at rest annihilation (allowing for a 511 keV line signal). This takes a time of the order of $\tau_{e}\sim 10^{3}-10^{6}$~years depending on the environmental density of electrons. We expect that the time scale until annihilation varies between the individual positrons, direction and energy of emission etc. by factors of at least ${\mathcal{O}}(1)$. In consequence a SN would contribute to the 511~keV scale for a time of the order of $\sim 10^{3}-10^{6}$~years. This is much longer than the typical time-interval between galactic SNe $\sim 50\,{\rm years}$. Hence, at any given time we receive signals from a sizable number of past SNe. This allows us to average over the Galactic distribution to get an estimate of the flux distribution. 

The probability distribution of type II SNe in the Galaxy is expected to follow the regions of high star formation, particularly the spiral arms. For this reason, it is peaked slightly off the Galactic center. However the exact distribution is still subject to some uncertainties~\cite{Mirizzi:2006xx,Ahlers:2009ae}. Here, we refer to the model presented in Ref.~\cite{Mirizzi:2006xx} for our quantitative analysis. We have verified that the more recent model presented in Ref.~\cite{Ahlers:2009ae} yields essentially identical results.

The probability distribution of Galactic SNe is best represented in the Galactocentric coordinate system, $(r,z,l)$, with the origin in the Galaxy center, the $x-$axis directed  toward the Sun, $r$ the radial coordinate in the Galactic plane, and $z$ the height, measured from the Galactic plane. The connection with the more common Galactic coordinate system can be made through the relations
\begin{eqnarray}
&& r = \sqrt{s^2 \cos^2 b + d_{\odot}^2 -2 d_\odot s\cos l \cos b}\,\ , \\ 
&& z=s \sin b \,,
\end{eqnarray}
where $- \pi \leq l \leq \pi$ is the Galactic longitude, $-\pi/2 \leq b \leq \pi/2$ is the Galactic latitude, $d_{\odot}=8.5$ kpc is the solar distance from the Galactic center and $s$ is the distance from the SN to the Sun.

The properly normalized SN volume distribution in the Galactocentric coordinate system is given by~\cite{Mirizzi:2006xx} 
\begin{equation}
n_{cc} = \sigma_{cc}(r) R_{cc}(z) \,\ \textrm{kpc}^{-3} \,\ , 
\label{eq:ncc}
\end{equation}
where $\sigma_{cc}(r)$ is the normalized Galactic surface density of type II events
\begin{equation}
\sigma_{cc}(r) = \frac{r^\zeta e^{-r/u}}{2\pi\,u^{2+\zeta}\,\Gamma(2+\zeta)} \,\ ,
\end{equation}
with $\Gamma$ the Euler gamma function. The vertical distribution, is approximated as a superposition of two Gaussian distributions with different scale height for the thin and thick disk,
\begin{eqnarray}
R_{cc}(z) &= &1.874\bigg\{  0.79 \exp\bigg[-\bigg(\frac{z}{0.212 \,\ \textrm{kpc}} \bigg)^2 \bigg] \nonumber \\
&+& 0.21   \exp\bigg[-\bigg(\frac{z}{0.636 \,\ \textrm{kpc}} \bigg)^2 \bigg]\bigg\} \,\ ,
\end{eqnarray}
Following Ref.~\cite{Mirizzi:2006xx}, in these expressions we take as benchmark values $\zeta =4$ and $u =1.25$ kpc.

\begin{figure*}
    \centering
    \begin{subfigure}
        \centering
        \includegraphics[width=\linewidth]{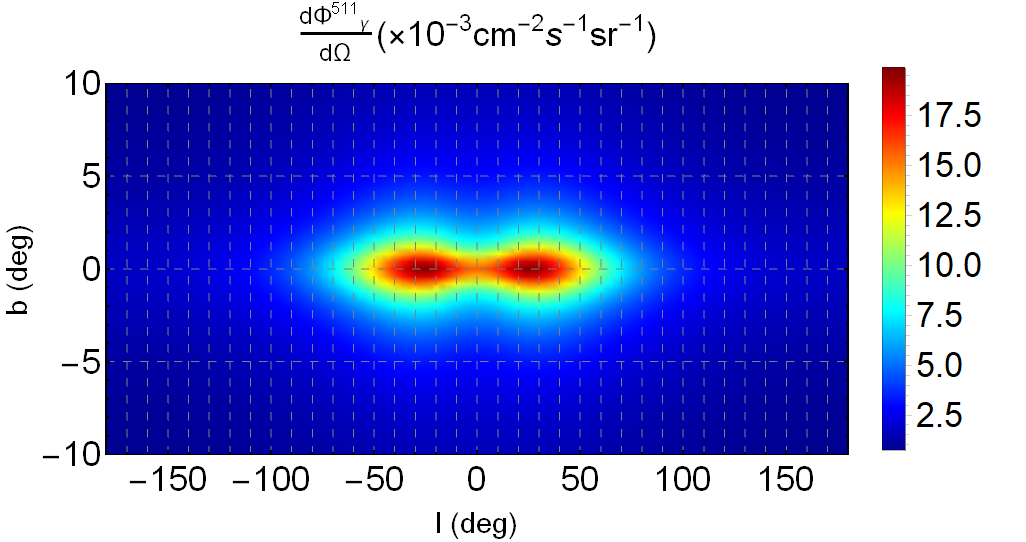} 
    \end{subfigure}
    \hfill
    \begin{subfigure}
        \centering
        \includegraphics[width=\linewidth]{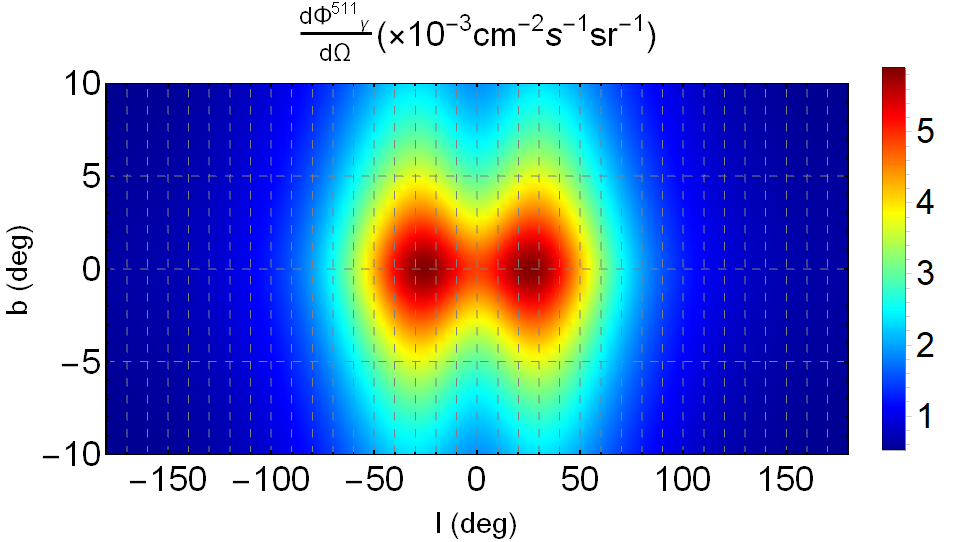} 
    \end{subfigure}
\caption{Sky-map in the region $b \in [-10^{\circ},10^{\circ}]$ and $l\in[-180^{\circ},180^{\circ}]$, of the photon flux in Eq.~\eqref{eq:photonflux} (upper panel), evaluated for $g_{ae}=3\times 10^{-12}$, $g_{ap}=10^{-9}$ and $m_a=30$~MeV.
In the lower panel we include a smearing at a scale of 1~kpc for $g_{ae}=1.5\times 10^{-18}$, $g_{ap}=10^{-9}$ and $m_a=30$~MeV. This simulates the effects of a non-negligible decay length of the ALP and the distance travelled by the positron before annihilating (see text for details).
}
\label{fig:skyph}
\end{figure*}

\begin{figure}[t!]
\vspace{0.cm}
\includegraphics[width=0.95\columnwidth]{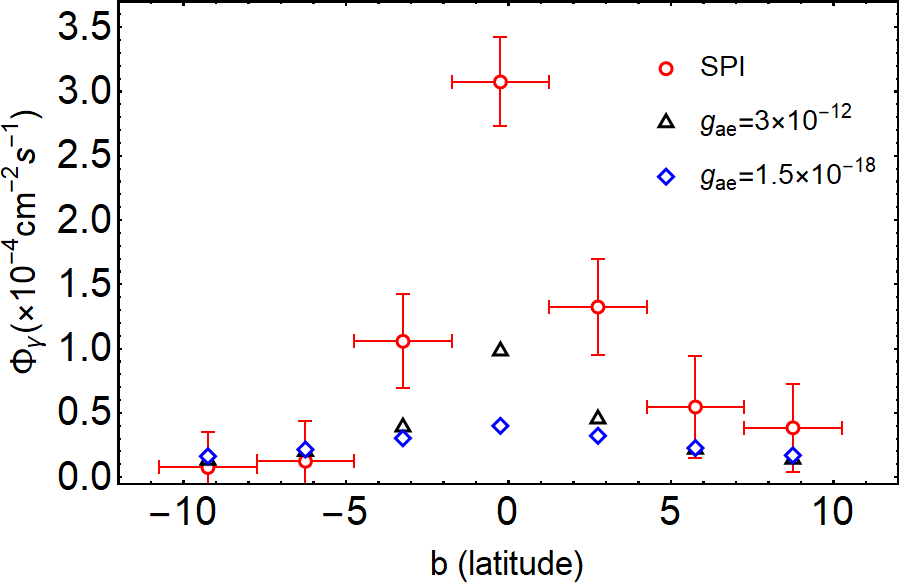}
\caption{Comparison of the photon flux produced by ALPs for $g_{ap}=10^{-9}$, $m_{a}=30$~MeV and two different values of $g_{ae}$ with the one measured by SPI \cite{Siegert:2019tus} as function of the Galactic latitude and integrated over the region $-5.25^{\circ}<l<3.75^{\circ}$. Note that we neglect a possible propagation of the positrons. For $g_{ae}=3\times 10^{-12}$ the smearing is done over $\lambda=l_e\ll 1$~kpc and has a negligible impact, while for $g_{ae}=1.5\times 10^{-18}$ we smeared the signal over 1 kpc, washing-out the central peak.
}
\label{fig:line2}
\vspace{0.3cm}

\includegraphics[width=0.95\columnwidth]{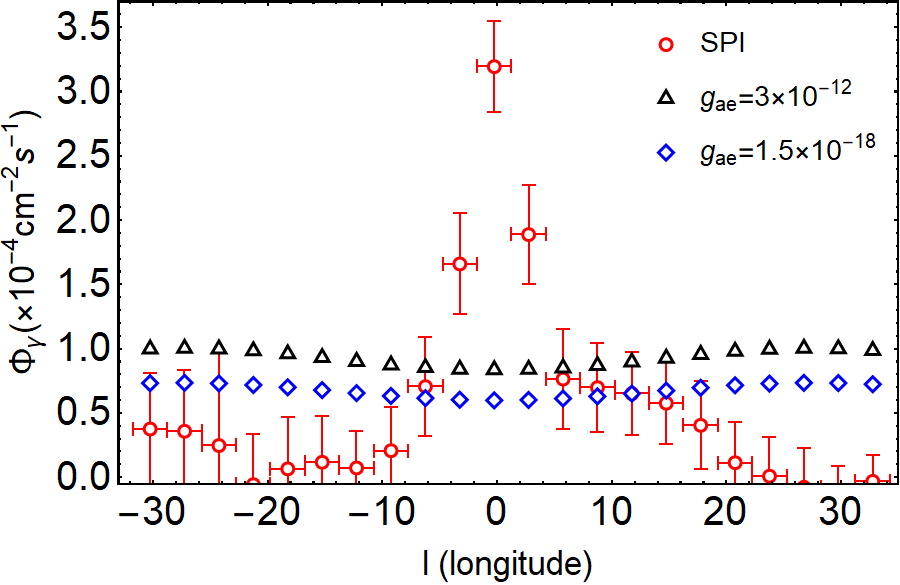}
\caption{As in Fig.~\ref{fig:line2} but as function of the Galactic longitude and integrated over the region $-10.75^{\circ}<b<10.25^{\circ}$. Note that the central dip is in agreement with Fig.~\ref{fig:skyph}. Due to the already much broader distribution of the SNe in this direction the smearing over 1~kpc for the smaller value of $g_{ae}$ is less pronounced.
}
\label{fig:line1}
\end{figure}
Then, the number of SNe exploded in a given infinitesimal element of volume per unit time is given by
\begin{equation}
\Gamma_{cc} n_{cc} \,\ s^2 ds \,\  d\Omega  \,\ ,
\end{equation}
where $d\Omega=d b \cos b\, dl $, and normalized such that $\int d\Omega \,ds\, s^{2}\,n_{cc}=1$. We fix the Galactic SN explosion rate to $\Gamma_{cc} =2$ SNe/century~\cite{Rozwadowska:2021lll}.\\
In our Galaxy, the emission of the 511 keV line proceeds mostly through the formation of positronium. Indeed, the positronium fraction is found to be $\sim$ 1 for positrons annihilating at rest in typical conditions of the Milky Way interstellar medium~\cite{Guessoum:2005cb}. Two photons of 511 keV energies originate from the (singlet) para-positronium state, leading to the specific line signal. The angular distribution in the Galactic sky-map of the 511 keV line photon signal produced by positron-electron annihilation through para-positronium formation is then given by:\footnote{Unless explicitly mentioned, here and in the following we neglect the small fraction of positrons ($\lesssim 25 \%$) that annihilate in flight and do not contribute to the signal. This has only a relatively small effect on the resulting limits.}
\begin{equation}
\frac{d \phi_\gamma^{ 511}}{d\Omega}= 2k_{ps}N_{\rm pos} \Gamma_{cc}\int ds \,  s^2 \frac{n_{cc}[r(s,b,l)]}{4 \pi s^2} \,\ ,
\label{eq:photonflux}
\end{equation}
where $k_{ps}=1/4$ accounts for the fraction of positrons annihilating through parapositronium,\footnote{Annihilation through orthopositronium leads to three photons that do not contribute to the 511~keV line signal.} then producing two photons with energy equal to 511~keV~\cite{Karshenboim:2003vs}.

In first approximation, the photon flux produced by ALP decays follows the Galactic SN distribution. The sky-map of the expected SN probability distribution and the resulting photon flux is shown in Fig.~\ref{fig:skyph} (upper panel). Under these conditions, we expect the positron distribution to be peaked at zero latitude, $b=0$, and at longitude $l \simeq \pm 30^\circ$, with a dip at $l=0$.

However, as discussed above, the distribution is smeared out to some degree by the combined effect of a significant decay length as well as by the distance travelled by the positrons before being stopped. By considering a longer ALP decay length for smaller values of the ALP mass or smaller couplings $g_{ae}$ and a distance $\sim 1$~kpc covered by positrons before annihilating, we expect the produced photon distribution to be smeared. We conservatively take this effect into account in two ways. Firstly, we use for the radius $r_G$ in Eq.~\eqref{eq:positrons} the smallest extend of the Galaxy, i.e. the vertical direction, $r_G=1\,{\rm kpc}$. Secondly, in order to account for the more diffuse emission, we smear the SN distribution over a scale $\lambda$, replacing $\sigma_{cc}(r)$ and $R_{cc}(z)$ in Eq.~\eqref{eq:ncc} respectively with
\begin{equation}\label{eq:smearing}
\begin{split}
 \sigma'_{cc}(r)&=A\,\int_0^{\infty} ds \sigma_{cc}(s)\,e^{-|s-r|/\lambda}\,,\\
R'_{cc}(z)&=B\,\int_{-\infty}^{\infty} ds R_{cc}(s)\,e^{-|s-z|/\lambda}\,,
\\
\end{split}
\end{equation}
where the normalization constants $A$ and $B$ are obtained by imposing $2\pi\int_0^\infty dr\,r\,\sigma_{cc}'(r)=1$ and $\int d\Omega\, ds\, s^2\, \sigma_{cc}'\,R_{cc}'=1$. The sky-map including this smearing is shown in Fig.~\ref{fig:skyph} (lower panel).

The expected 511 keV photon flux from ALP decays should be compared with the 511 keV flux measurement from SPI. The 511 keV photon flux $\Phi_\gamma$ distributions in latitude $b$ and in longitude $l$, as derived in Ref.~\cite{Siegert:2019tus}, are shown in Figs.~\ref{fig:line2}-\ref{fig:line1} respectively, where we overlay the ALP-induced signal for $m_a=30$~MeV and two representative values of the coupling, $g_{ae}= 3\times 10^{-12}$ and $g_{ae}= 1.5\times 10^{-18}$. To demonstrate the impact of the smearing, we blurred the signal over a scale $\lambda=\min(l_e,\,1\,\kpc)$ following the recipe in Eq.\eqref{eq:smearing}, ignoring the possible positron propagation before their annihilation. As expected, the photon flux produced by ALP decay is peaked at $b=0$. For larger electron couplings, the shape of the ALP signal becomes closer to the observed photon flux in the latitude direction  (Fig.~\ref{fig:line2}). However, as shown in Fig.~\ref{fig:line1}, the longitudinal distribution of the photons from ALPs is very different from the 511 keV signal observed by the  SPI. Indeed, as discussed above, there is a dip at $l=0$ where the observed signal is strongly peaked. Moreover, the two peaks at $l \simeq \pm 30^\circ$ are clearly visible.

In order to get a conservative bound on the ALP parameter space for our analysis we consider the longitude and latitude profiles from Ref.~\cite{Siegert:2019tus}. Given the uncertainties in our calculation, we use a very simple procedure to set our limit: the bound on ALPs is set by the first bin where the predicted signal exceeds the data at 2$\sigma$. In our case, the most constraining bin comes from the longitudinal distribution at $l\in[28.25^{\circ},31.25^{\circ}]$ because the predicted photon distribution is almost flat in this direction.

\begin{figure}[t!]
\vspace{0.cm}
\includegraphics[width=0.95\columnwidth]{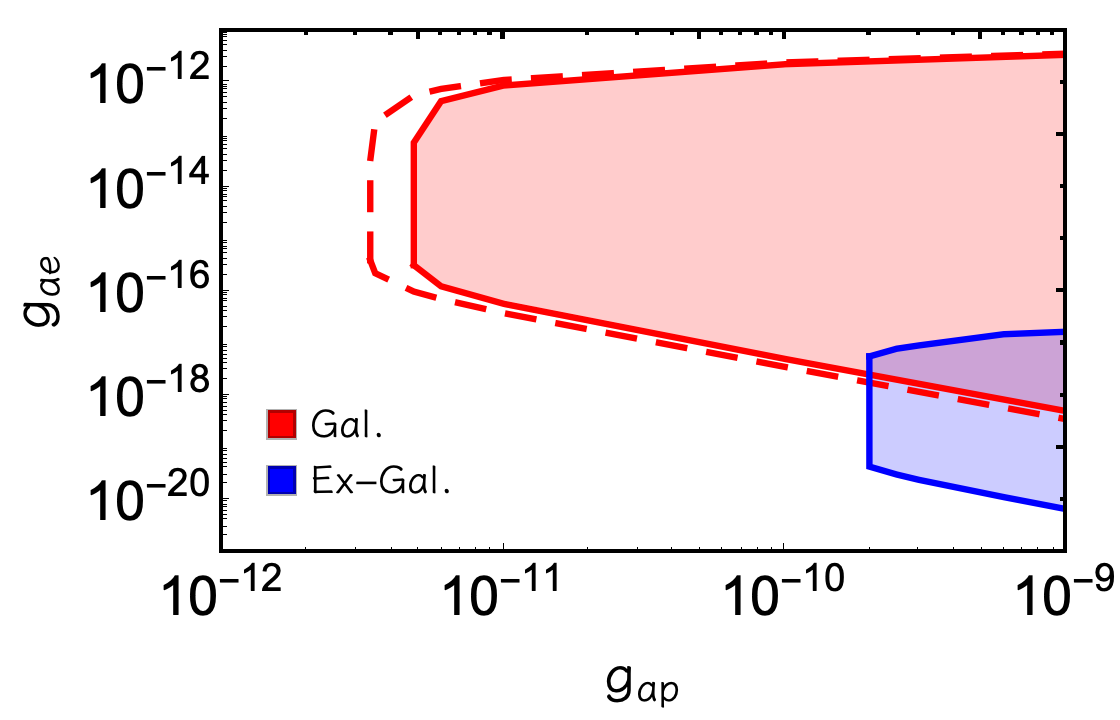}
\caption{Bounds on $g_{ae}$ vs $g_{ap}$ from the diffuse Galactic SNe (red band) and 
from extra-galactic SNe (blue band) for $m_a=30$~MeV. The dashed red line illustrates the effect of smearing by assuming the extreme case where the photon signal follows exactly the distribution of SNe, i.e. it neglects the distance travelled by ALPs and positrons. Values $g_{ap}\gtrsim 10^{-9}$ are excluded by the energy loss argument~\cite{Burrows:1988ah,Keil:1996ju,Raffelt:2006cw,Carenza:2019pxu}.}
\label{fig:bound}
\end{figure}

In Fig.~\ref{fig:bound}, we plot the bound (red region) on the ALP-electron coupling $g_{ae}$ as a function of the ALP-nucleon coupling $g_{ap}$ in the range below the SN 1987A energy-loss bound. In order to be conservative, our fiducial exclusion region (the shaded red region) is obtained smearing the ALP-induced photon signal over a scale $\lambda=1$~kpc, taking into account both the ALP propagation before their decay and the positron one before annihilation. For comparison, the red dashed line represents a more optimistic bound, where we assume that ALPs decay and positrons annihilate in the immediate vicinity of where they are produced. The boundaries of the excluded band $g_{ae}^L \lesssim g_{ae} \lesssim g_{ae}^H$ are given by the following arguments. For $g_{ae} \gtrsim  g_{ae}^H$, ALPs would decay inside the SN envelope. Analogously, for $g_{ae} \lesssim g_{ae}^L$,  ALPs would escape our Galaxy before decaying into pairs. In both cases, such ALP decays could not contribute to the 511 keV signal. We see that for $g_{ap}\sim 10^{-9}$, the range $10^{-19} \lesssim g_{ae} \lesssim 10^{-12}$ is excluded. 

\begin{figure}[t!]
\vspace{0.cm}
\includegraphics[width=0.95\columnwidth]{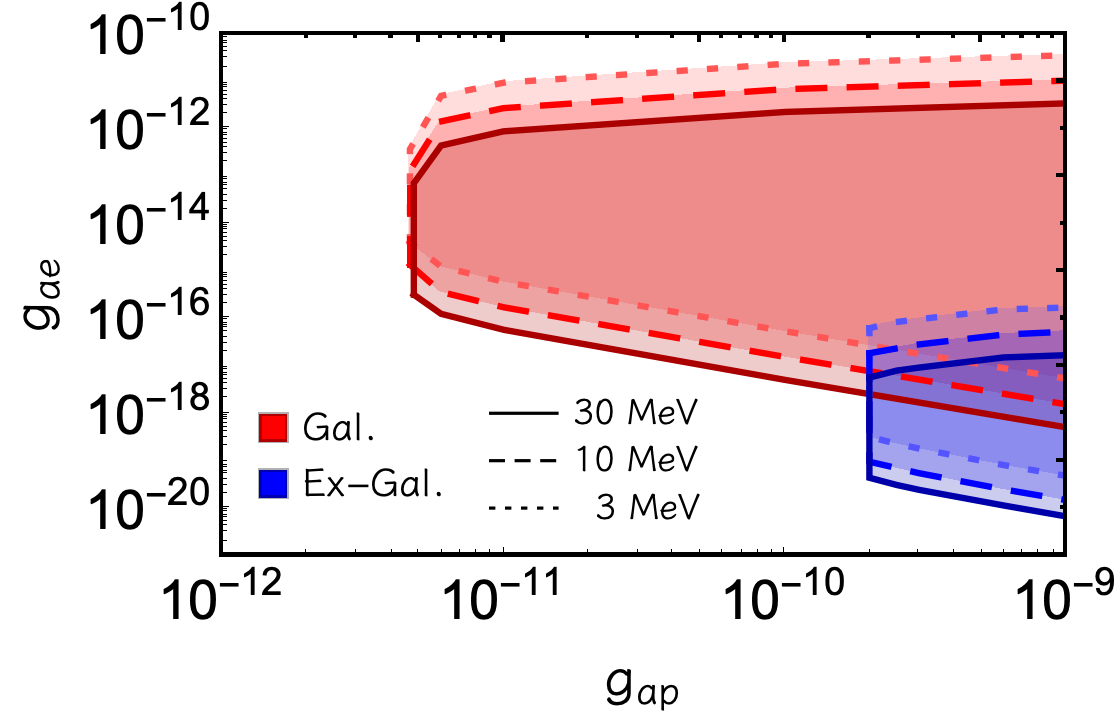}
\caption{Bounds on $g_{ae}$ vs $g_{ap}$ from the diffuse Galactic SNe (reddish bands)  and from extra-galactic SNe (bluish bands) for three representative values of the ALP mass $m_a= 3,\,10,\,30$~MeV.}
\label{fig:boundmass}
\end{figure}

For smaller values of $g_{ap}$, the upper boundary of the exclusion plot is rather flat, while the lower boundary is reduced becoming $g_{ae} \gtrsim 10^{-17}$ for $g_{ap}\sim 10^{-11}$. This different behaviour is due to the fact that  along the upper boundary the positron flux would exceed the 511 keV signal independently of the exact value of $g_{ap}$ in the chosen range, as long as the ALPs decay outside the SN envelope. The latter condition fixes the upper value of $g_{ae}$. Conversely, along the lower boundary, since $g_{ae}$ is much smaller, a sizable value of $g_{ap}$ is required in order to get an observable signal. 

Reducing the value of $g_{ap}$, we reach a point at which ALPs are not abundant enough to produce a detectable signal even though they decay entirely into positrons. This threshold is represented in Fig.~\ref{fig:bound} by the vertical left end of the exclusion region.

We note that for $g_{ae}\gtrsim 10^{-12}$ the ALP decay length is comparable with the SN photosphere radius, $l_{e}\sim r_{\rm esc}$. As discussed in Ref.~\cite{DeRocco:2019njg}, in this situation ALPs might create a \emph{fireball}, i.e. a layer of hot electron-positron plasma due to the ALP decays just outside the photosphere (located at $r_{\rm esc}$), which absorbs the positrons produced in the ALP decay. However, the total number of positrons escaping the SN is only an order-one factor smaller than the initial number produced at  $r_{\rm esc}$. Therefore, the formation of a fireball has a small effect on the positron flux and thus our bound is mostly unaffected by this phenomenon.

As shown in Fig.~\ref{fig:boundmass}, the shape of the exclusion region is rather independent of the ALP mass. The dominant effect of the mass is that the bound is shifted towards larger couplings as the mass decreases. This is due to the dependence of the ALP decay length on the mass.

We finally note that our bound, based on the angular structure of the ALP-induced signal, might be translated into the requirement that a single SN should emit no more than $\sim 10^{52}$ positrons, a value one order of magnitude more stringent than the estimate of Ref.~\cite{DeRocco:2019njg}. More precisely, without smearing a SN cannot emit more than $N_{\rm pos}\lesssim 8\times 10^{51}$, while with the smearing we find $N_{\rm pos}\lesssim 1.6\times 10^{52}$.\footnote{In these numbers we have accounted for the positrons annihilating in flight by conservatively assuming that this fraction is at most 25\% in the relevant energy range.}

\bigskip

In addition to the combined flux from the Galactic SNe, one may also look for a 511 keV signal from young local SNe and SN remnants, in order to derive constraints on positron emission from different sources~\cite{Martin:2010hw}. Among the sources observed by SPI, the only one corresponding to a Galactic type II SN explosion (SNIIb) is the SN remnant Cassiopea A, at a distance $d=3.4$ kpc~\cite{1995ApJ...440..706R}, which exploded about $t_{\rm CAS}=400$ years ago~\cite{2001AJ....122..297T}. However, since the explosion time of this source is much smaller than the typical positron annihilation time in the Galaxy,\footnote{As the slow down requires many interactions, we think that the number of slowed down positrons does not simply follow an exponential decay law in their ``lifetime''. Instead we expect it to be more peaked around the lifetime.} it is uncertain whether we receive a sizable contribution to the 511 keV signal from ALP emission. 

\subsection{Extra-galactic supernovae}
\label{sec:extra-galactic}

We now consider the case of extra-galactic SNe, producing ALPs that decay into positrons outside our Galaxy. In the extra-galactic medium, where charged particles are trapped by $B\sim \mathcal{O}(1)$~nG, the produced relativistic positrons would slow down and annihilate at rest on a timescale comparable or faster than the Hubble expansion time~\cite{Iguaz:2021irx} ($\tau_e \lesssim 10^{10}$ yrs), giving two photons, each with energy $m_e$, as in the previous case. However, such photons are redshifted and thus do not contribute to the 511~keV line.
Rather, they contribute to the cosmic X-ray background (CXB), measured by different experiments, e.g. the High Energy Astronomy Observatory (HEAO)~\cite{McHardy:1997fb} and the Solar Maximum Mission (SMM) \cite{doi:10.1063/1.53933} (a recent data compilation can be found in Ref.~\cite{Ballesteros:2019exr}). This case is useful to extend the previous constraints to smaller values of $g_{ae}$. The cumulative energy flux of escaping ALPs from past type II SNe, and decaying into electron-positron pairs in the redshift interval between $[z_d:z_d-dz_d]$  is given by (see, e.g., Refs.~\cite{Raffelt:2011ft,Calore:2020tjw})
\begin{eqnarray}
& &\!\!\!\!\!\!\!\!\!\!\!\!\!\!\!\!\!\! \left(\frac{d\phi_{a}(E_{a})}{dE_{a}} \right)_{\rm dec}
\\\nonumber
&=& \int_{z_{d}}^{\infty}dz\,(1+z)\frac{dN_{a}(E_{a}(1+z))}{dE_{a}}R_{SN}(z)\left|\frac{dt}{dz}\right|
\\\nonumber 
& &\qquad\qquad\times 
\left[e^{-(z-z_{d})/H_{0} l_e}-e^{-(z-z_{d}+dz_{d})/H_{0} l_e}\right] \,\,
\end{eqnarray}
where $z$ is the redshift. $R_{SN}(z)$ is the SN explosion rate, taken from~\cite{Priya:2017bmm}, with a total normalization for the type II rate $R_{cc}=1.25 \times 10^{-4}$yr$^{-1}$ Mpc$^{-3}$. Furthermore, $ |{dt}/{dz} |^{-1}= H_0(1+z)[\Omega_\Lambda+\Omega_M(1+z)^3]^{1/2}$ with the cosmological parameters $H_0= 67.4$ km s$^{-1}$ Mpc$^{-1}$, $\Omega_M=0.315$, $\Omega_\Lambda=0.685$~\cite{Aghanim:2018eyx}. Most of the contribution to the ALP flux comes from $z \sim 1-2$.

Expanding the previous expression for small $d z_d$ one finds the differential flux of decayed ALPs,
\begin{eqnarray}
& &\!\!\!\!\!\!\!\!\!\!\!\!\!\! \bigg(\frac{d^2 \phi_a (E_a)}{d E_a d z_d}\bigg)_{\rm dec} \\\nonumber
&=& \int_{z_d}^{\infty} \! (1+z) \frac{dN_a(E_a(1+z))}{dE_a}
\\\nonumber 
&&\qquad\times 
[R_{SN}(z)] \exp\bigg(-\frac{z-z_d}{H_0 l_e} \bigg) \frac{1}{H_0 l_{e}} \bigg[ \bigg|\frac{dt}{dz} \bigg| dz \bigg] \,.
\label{eq:diffuse}
\end{eqnarray}
Since the major contribution to the ALP flux comes from $z\lesssim2$, the photons produced by the annihilation of the positrons, originated from ALP decays, are not absorbed~\cite{Arcodia:2018sct} and, due to the redshift, reach us with an energy 
\begin{equation}
E_{\gamma}= \frac{m_e}{1+z_d} \,\ .
\end{equation}
The produced photon flux is then given by
\begin{eqnarray}
\frac{d \phi_\gamma}{d E_\gamma}
&=&2 k_{ps} \frac{d \phi_a}{d z_d} 
\frac{d z_d}{d E_\gamma} \nonumber \\
&=&2 k_{ps}
\frac{m_e}{E_\gamma^2}\int_{m_a}^{\infty} dE_a  \bigg(\frac{d^2 \phi_a (E_a)}{d E_a d z_d}\bigg)_{\rm dec} \,\ .
\end{eqnarray}

\begin{figure}[t!]
\vspace{0.cm}
\includegraphics[width=0.95\columnwidth]{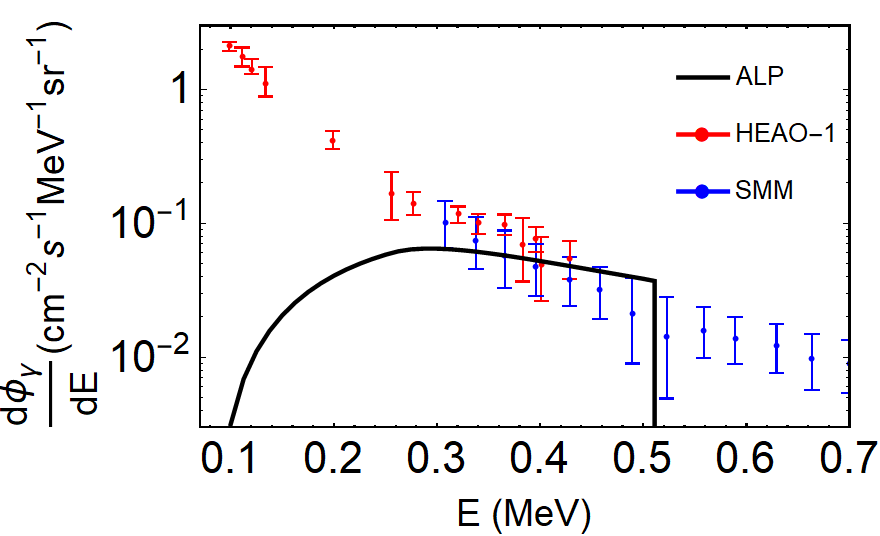}
\caption{Photon flux from CXB measured by HEAO-1~\cite{McHardy:1997fb} and SMM~\cite{doi:10.1063/1.53933} with $2\sigma$ error bars, compared with the X-ray flux from ALP decays from extra-galactic SNe for $m_a=30$~MeV, $g_{ap}=10^{-9}$, $g_{ae}=7 \times 10^{-21}$ (solid curve).}
\label{fig:cxb}
\end{figure}

In Fig.~\ref{fig:cxb}, we compare our result for the photon flux produced by ALP decays from extra-galactic SNe (assuming $m_a=30$~MeV, $g_{ap}=10^{-9}$, $g_{ae}=7 \times 10^{-21}$) 
to the CXB flux measured by HEAO-1~\cite{McHardy:1997fb} and SMM~\cite{doi:10.1063/1.53933}. 
%\sout{We compare it with the photon flux produced ALP decays from extra-galactic SNe for $m_a=30$~MeV, $g_{ap}=10^{-9}$, $g_{ae}=7 \times 10^{-21}$.} 
In order to get a bound on $g_{ae}$ vs $g_{ap}$, we require that the produced photon flux from ALP decays does not exceed the measured CXB by more than 2$\sigma$. We show the exclusion areas also in Fig.~\ref{fig:bound}, for $m_{a}=30$~MeV (blue region), and in Fig.~\ref{fig:boundmass}, where we compare three values of the ALP mass $m_a=3,\,10,\,30$~MeV (light blue regions).  
 
In this case, the upper limit is obtained by requiring that the decayed photons are in the X-ray band accessible to the instruments used for detection, while the lower limit is given by the requirement that ALPs decay before reaching our Galaxy. At $g_{ap}=10^{-9}$, the lower bound is one order of magnitude more stringent than the bound from Galactic SNe for the corresponding ALP mass. In particular, values $g_{ae}\gtrsim10^{-20}$ are excluded for $g_{ap}=10^{-9}$ and $m_a=30$~MeV. The trend of $g_{ae}$ vs $g_{ap}$ is similar to what observed in the Galactic case. However, we can lower the value of $g_{ap}$ only by less than one order of magnitude before the bound disappears.

We remark that our bounds from the CXB are conservative in the sense that we do not attempt any modeling and subtraction of the guaranteed astrophysical contributions from  extra-galactic objects. Indeed, models from active galactic nuclei quite successfully explain the whole CXB up to at least 200 keV~\cite{Ueda:2014tma}, while at higher energies -- which are the ones most relevant for our bound -- their contribution remains more uncertain.

\section{Discussion and Conclusions}
\label{sec:conclusions}

In this paper we have investigated the physics potential of Galactic and extra-galactic type II SNe to constrain ALPs coupled with nucleons and electrons.

In such a situation, ALPs are produced in the SN core via nucleon-nucleon bremsstrahlung. Their decay then produces electron-positron pairs. The positrons are stopped by interactions with matter and then annihilate with electrons to produce 511~keV photons. For Galactic SNe this produces a 511 keV gamma-ray line. Using observations of the spectrometer SPI (SPectrometer on INTEGRAL)~\cite{Siegert:2019tus}, we obtain stringent constraints for the electron-ALP coupling, excluding the range $10^{-19} \lesssim g_{ae} \lesssim 10^{-12}$ for $g_{ap}\sim 10^{-9}$. ALPs from extra-galactic SNe are stopped (on significantly longer length/time-scales) by the extra-galactic medium. In this case, the red-shift has to be taken into account leading to a somewhat broader signal. Nevertheless, data from the observation of the cosmic X-ray background~\cite{Ballesteros:2019exr} improve the previous constraint down to $g_{ae} \sim 10^{-20}$. Further improvement could result from observations of the future
eASTROGAM~\cite{DeAngelis:2017gra} and AMEGO~\cite{McEnery:2019tcm}. In particular additional measurements away from the Galactic center region would be helpful. Also, improvements in the understanding and detailed modeling of the positron propagation and slow-down would strengthen the confidence in the results, but potentially also allow to tighten the limits via a better modeling of the morphology. To our knowledge, the discussed range of couplings is currently unexplored. Moreover, as briefly discussed in Sec.~\ref{sec:production}, the constrained range of couplings is easily motivated in simple models where the ALPs are pseudo-Goldstone bosons. 

Beyond our result on ALPs, we also note that we obtain an improved result on the maximum number of positrons that may be emitted by an SN outside its envelope $N_{\rm pos}\lesssim 10^{52}$, which takes into account the SN distribution inside the Galaxy. This result could also be applied to update limits on other particles such as dark photons~\cite{DeRocco:2019njg}.

\section*{Acknowledgments}
We warmly acknowledge enlightening discussions with T. Siegert, and we also thank him for having provided us with data from~\cite{Siegert:2019tus} in a tabulated form.
The work of P.C., G.L. and A.M. is partially supported by the Italian Istituto Nazionale di Fisica Nucleare (INFN) through the ``Theoretical Astroparticle Physics'' project and by the research grant number 2017W4HA7S ``NAT-NET: Neutrino and Astroparticle Theory Network'' under the program PRIN 2017 funded by the Italian Ministero dell'Universit\`a e della Ricerca (MUR).The work of F.C. is partially supported by the ``Agence Nationale de la Recherche'', grant n.~ANR-19-CE31-0005-01 (PI: F. Calore). 

\section*{Appendix: Quantifying the uncertainties of the model}
\label{sec:app}
\begin{figure}[t!]
\vspace{0.cm}
\includegraphics[width=0.95\columnwidth]{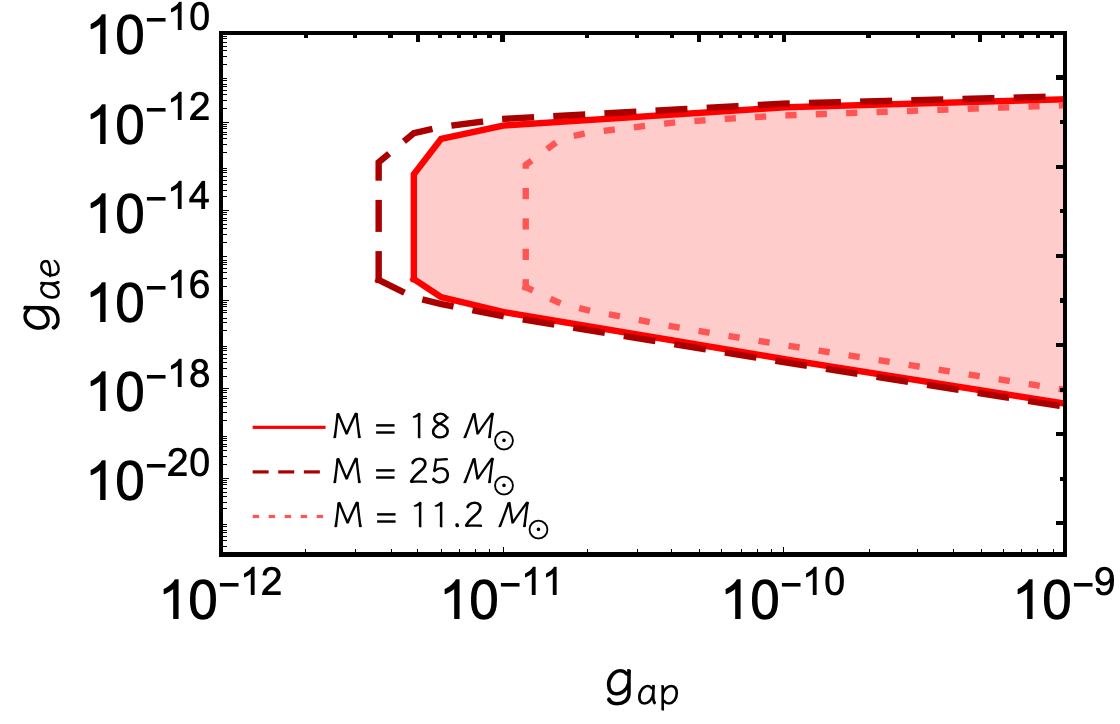}
\caption{Galactic bound on $g_{ae}$ vs $g_{ap}$ for $m_a=30$~MeV using three SN models with different progenitor masses: $M=18~M_\odot$ (continuous red line), $M=25~M_\odot$ (dashed darker red) and $M=11.2~M_\odot$ (dotted red line).}
\label{fig:progenitors}
\end{figure}
In this Appendix we discuss the role of the two major sources of uncertainties encountered in this work: the dependence of the ALP flux from the SN progenitor and the choice of the smearing scale $\lambda$ in Eq.~\eqref{eq:positrons}.\\ 
The first is quantified by considering two other different SN models, with $25~M_\odot$ and $11.2~M_\odot$ progenitor masses. We observe that the SN temperature grows as the progenitor mass increases, therefore at fixed value of $g_{ap}$ the produced ALP flux is larger and the typical energy is higher for heavier progenitors. In addition, we checked that the assumption of massless ALPs for $m_a\lesssim 30$~MeV is still valid for the lighter progenitor. For the $25~M_\odot$ model, the best-fit parameters of the ALP flux in Eq.~\eqref{eq:time-int-spec} are $C=1.1\times 10^{56}$ ${\rm MeV}^{-1}$, $E_0=126.8$ MeV and $\beta=2.03$ for the reference couplings $g^{\rm ref}_{ap}=10^{-9}$ and $g_{an}=0$. On the other hand, the ALP flux for the $11.2~M_\odot$ model is fitted by $C=2.81\times 10^{55}$ ${\rm MeV}^{-1}$, $E_0=79.5$ MeV and $\beta=2.5$ for the same reference couplings as before. The $11.2~M_{\odot}$ progenitor is a motivated representative SN model because the SN population is larger at lower masses~\cite{future}. In Fig.~\ref{fig:progenitors} we show the Galactic bound evaluated for the three different representative SN progenitors. As expected, the exclusion region becomes smaller as the progenitor mass decreases, due to the reduction of the produced ALP flux.
\\
\begin{figure}[t!]
\vspace{0.cm}
\includegraphics[width=0.95\columnwidth]{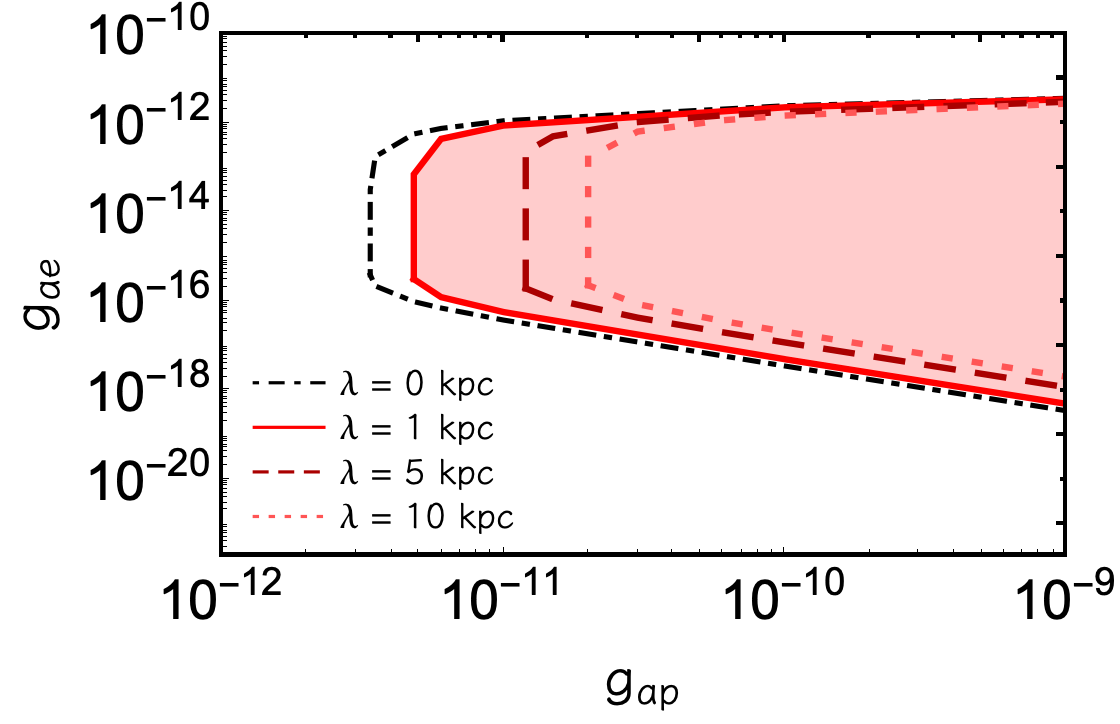}
\caption{Galactic bound on $g_{ae}$ vs $g_{ap}$ for $m_a=30$~MeV using different smearing scales $\lambda$, as shown in legend. The dot-dashed black line is the bound obtained without smearing, neglecting the distance travelled by the ALPs and positrons as in Fig.~\ref{fig:bound}.
}
\label{fig:diff_sm}
\end{figure}
The uncertainty related to the smearing scale $\lambda$ is addressed by calculating the bound for two additional values of this scale, namely $5$~kpc and $10$~kpc. In Fig.~\ref{fig:diff_sm} we compare the case without smearing (dot-dashed black curve), our fiducial bound with $\lambda=1$~kpc (the shaded red area in the solid red line) and the other two cases with $\lambda=5$~kpc (dashed dark red line) and $\lambda=10$~kpc (dotted light red line). The bound is weakened as the smearing scale increases. Indeed, for larger values of $\lambda$ the signal becomes more featureless and the equivalent number of produced positrons from each SN tends to increase. More precisely, accounting for the positrons annihilating in flight as discussed in Sec.~\ref{sec:positron}, for $\lambda=5$~kpc we obtain $N_{\rm pos}\lesssim8.7\times 10^{52}$ and for $\lambda=10$~kpc we obtain $N_{\rm pos}\lesssim2.6\times 10^{53}$.
% To summarize,  
% the largest uncertainty is given by a proper inclusion of the positron propagation in the Galaxy, a topic that will be considered for a future and more detailed study.
Clearly, a more accurate description of the positron propagation in the Galaxy would reduce the uncertainties in our analysis. 
%This investigation is postponed to a future and more detailed study. 

\bibliographystyle{utphys}
\bibliography{references.bib}

\end{document}